\documentclass[12pt]{article}
\usepackage{times}
\usepackage{marvosym}
\usepackage{geometry}
\usepackage{endnotes}
\usepackage[labelfont=bf]{caption}
\usepackage{inputenc,graphicx}
\usepackage[most]{tcolorbox}
\usepackage{enumitem,amssymb}
\usepackage{ragged2e}
\usepackage{xcolor}
\newlist{thematic}{itemize}{8}
\setlist[thematic]{label=$\square$}
\usepackage{pifont}
\newcommand{\xmark}{\ding{55}}%

\usepackage{hyperref}

\makeatletter
\def\enoteheading{\section*{\notesname
  \@mkboth{\MakeUppercase{\notesname}}{\MakeUppercase{\notesname}}}%
  \mbox{}\par\vskip-2.3\baselineskip\noindent\rule{.5\textwidth}{0.4pt}\par\vskip\baselineskip}
\makeatother

\usepackage{ulem}
\let\footnote=\endnote

\begin{document}
\raggedright
\huge
Astro2020 APC White Paper \linebreak

Primarily Undergraduate 
Institutions and the Astronomy Community   \linebreak
\normalsize

\noindent \textbf{Thematic Areas:} \hspace*{60pt} $\square$ Ground Based Project \hspace*{10pt} $\square$ Space Based Project \hspace*{20pt}\linebreak
$\square$ Infrastructure Activity \hspace*{31pt} $\square$ Technological Development Activity \linebreak
  \text{\rlap{\xmark}}$\square$ State of the Profession Consideration \hspace*{65pt} \linebreak
  
\textbf{Principal Author:}

Name: Joseph Ribaudo
 \linebreak						
Institution: Providence College, Utica College
 \linebreak
Email: \texttt{jribaudo@providence.edu}
 \linebreak
Phone:  401.865.2379
 \linebreak
 
\textbf{Co-Authors:} (names, institutions, email) \linebreak
Rebecca A. Koopmann, Union College, \texttt{koopmanr@union.edu}\linebreak
Aileen A. O'Donoghue, St. Lawrence University, \texttt{aodonoghue@stlawu.edu}\linebreak
Aparna Venkatesan, University of San Francisco, \texttt{avenkatesan@usfca.edu} \linebreak

\pagenumbering{gobble}

\textbf{Co-Signers:} (names, institutions, email) \linebreak
Thomas J. Balonek, Colgate University, \texttt{tbalonek@colgate.edu} \linebreak
Rachael L. Beaton, Princeton/Carnegie Observatories, \texttt{rbeaton@princeton.edu} \linebreak
Jillian Bellovary, CUNY - Queensborough Community College, \texttt{jbellovary@amnh.org} \linebreak
G. Bruce Berriman, Caltech/IPAC-NExScI, \texttt{gbb@ipac.caltech.edu}\linebreak
John J. Bochanski, Rider University, \texttt{jbochanski@rider.edu}\linebreak
Derek L. Buzasi, Florida Gulf Coast University, \texttt{dbuzasi@fgcu.edu}\linebreak
John M. Cannon, Macalester College, \texttt{jcannon@macalester.edu} \linebreak
Joleen K. Carlberg, Space Telescope Science Institute, \texttt{jcarlberg@stsci.edu}\linebreak
Jennifer L. Carter, Susquehanna University, \texttt{carterj@susqu.edu}\linebreak
Charlotte Christensen, Grinnell College, \texttt{christenc@grinnell.edu} \linebreak
Kim Coble, San Francisco State University, \texttt{kcoble@sfsu.edu}\linebreak
Kevin R. Covey, Western Washington University, \texttt{coveyk@wwu.edu}\linebreak
Mary Crone Odekon, Skidmore College, \texttt{mcrone@skidmore.edu} \linebreak
Kathryn Devine, The College of Idaho, \texttt{kdevine@collegeofidaho.edu} \linebreak
Chuanfei Dong, Princeton University, \texttt{dcfy@princeton.edu} \linebreak
Adriana Durbala, University of Wisconsin-Stevens Point, \texttt{adurbala@uwsp.edu} \linebreak
Paul B. Eskridge, Minnesota State University, Mankato, \texttt{paul.eskridge@mnsu.edu} \linebreak
Cassandra Fallscheer, Central Washington University, \texttt{cassandra.fallscheer@cwu.edu} \linebreak
Gregory A. Feiden, University of North Georgia, \texttt{gregory.feiden@ung.edu} \linebreak
Rose A. Finn, Siena College, \texttt{rfinn@siena.edu} \linebreak
Josh T. Fuchs, Texas Lutheran University, \texttt{jfuchs@tlu.edu} \linebreak 
Gregory Hallenbeck, Washington \& Jefferson College, \texttt{ghallenbeck@washjeff.edu}\linebreak
Martha Haynes, Cornell University, \texttt{haynes@astro.cornell.edu}\linebreak
Todd Hillwig, Valparaiso University, \texttt{Todd.Hillwig@valpo.edu}\linebreak
Cameron Hummels, California Institute of Technology, \texttt{chummels@caltech.edu} \linebreak
Richard R. Lane, Pontificia Universidad Cat\'olica de Chile, \texttt{rlane@astro.puc.cl}\linebreak
Lauranne Lanz, Dartmouth College/The College of New Jersey,
\texttt{lauranne.lanz@dartmouth.edu}\linebreak
Nicolas Lehner, University of Notre Dame, \texttt{nlehner@nd.edu}\linebreak
Lukas Leisman, Valparaiso University, \texttt{luke.leisman@valpo.edu}\linebreak
Britt Lundgren, University of North Carolina Asheville, \texttt{blundgre@unca.edu}\linebreak
Karen Masters, Haverford College, \texttt{klmasters@haverford.edu}\linebreak
John O'Meara, W.M. Keck Observatory, \texttt{jomeara@keck.hawaii.edu} \linebreak
D.J. Pisano, West Virginia University, \texttt{djpisano@mail.wvu.edu} \linebreak
Katie Rabidoux, University of Wisconsin-Platteville, \texttt{rabidouxk@uwplatt.edu} \linebreak
Benjamin Rose, Space Telescope Science Institute, \texttt{brose@stsci.edu}\linebreak
Jessica Rosenberg, George Mason University, \texttt{jrosenb4@gmu.edu} \linebreak
Michael J. Rutkowski, Minnesota State University-Minnesota, \texttt{michael.rutkowski@mnsu.edu} \linebreak
J. Allyn Smith, Austin Peay State University, \texttt{smithj@apsu.edu} \linebreak
Ryan Terrien, Carleton College, \texttt{rterrien@carleton.edu}\linebreak
Todd Tripp, University of Massachusetts-Amherst, \texttt{tripp@astro.umass.edu}\linebreak
Parker Troischt, Hartwick College, \texttt{TroischtP@hartwick.edu} \linebreak
Nicholas Troup, Salisbury University, \texttt{nwtroup@salisbury.edu}\linebreak
Jackie Villadsen, St.~Mary's College of Maryland, \texttt{jrvilladsen@smcm.edu} \linebreak
Anna Williams, Macalester College, \texttt{awilli11@macalester.edu}\linebreak
Jason E. Ybarra, Bridgewater College, \texttt{jybarra@bridgewater.edu}\linebreak

\justifying

\pagebreak
\pagenumbering{arabic}
\section*{Executive Summary}

This White Paper highlights the role Primarily Undergraduate Institutions (PUIs) play within the astronomy profession, addressing issues related to employment, resources and support, research opportunities and productivity, and educational and societal impacts, among others. Astronomers
working at PUIs are passionate about teaching and mentoring undergraduate students through 
substantive astronomy experiences, all while working to continue research programs that 
contribute to the advancement of the professional field of astronomy. PUIs are where the majority of undergraduate students pursue post-secondary education, and as such, understanding the unique challenges and opportunities associated with PUIs is critical to fostering an inclusive astronomy community throughout the next decade.

We provide a view of the profession as lived and experienced by faculty and students of PUIs, while highlighting the unique opportunities, challenges, and obstacles routinely faced.  A variety of recommendations are outlined to provide the supporting structures and resources needed for astronomy to thrive at PUIs over the next decade and beyond - a critical step for a profession focused on fostering and maintaining an inclusive, supportive, and diverse community. \vspace{0.25 cm}

\begin{tcolorbox}[left=0mm,right=2mm,width=6.5in,colback=gray!5!white,colframe=gray!75!black]
\textbf{\large Recommendations:}
\justify
\begin{itemize}[leftmargin=*]
    \item \textbf{The Decadal Survey:} Address the significant under-representation of 
    PUI faculty membership on Decadal panels and committees. Funding course releases 
    for PUI faculty may be one way to improve participation.
    \item \textbf{Teaching and Pedagogy:} Provide greater opportunities for teaching 
    courses and teacher-training/mentoring for graduate students who aspire to faculty 
    positions at PUIs (similar to the research mentoring of a Ph.D. program).
    \item \textbf{Research Support:} Increase the number of targeted grant programs for 
    faculty at PUIs. Consider extending these PUI-awards for up to 5 years, 
    to allow for sustainable projects with undergraduate student research participation. 
    Foster programs that allow PUI faculty to interact with or mentor graduate
    students and postdoctoral researchers.
    \item \textbf{PUI-focused Collaborations:} Encourage and financially support the 
    development of large-scale, long-duration research collaborations that leverage the contributions of PUI-affiliates (faculty and students). See the White Paper by 
    Koopmann et al. - \textit{Integrating Undergraduate 
    Research and Faculty Development in a Legacy Astronomy Research Project}.
    \item \textbf{Regional Astronomy:} Increase the number of regional astronomy 
    associations and provide financial and logistical support for the establishment of 
    regularly held regional meetings.
    \item \textbf{Scholarship and Career Assessment:} Establish a robust assessment 
    procedure to examine the scholarly research contributions made by PUI-affiliates
    (publications and conference presentations). Expand current 
    career assessment work to include an analysis of astronomers at PUIs. Create
    formal AAS committees/working groups to pursue PUI-related assessment.
\end{itemize}
\end{tcolorbox}


\pagebreak
\section{Background}\vspace{-0.25 cm}

Primarily Undergraduate Institutions (PUIs) are defined by the National
Science Foundation (NSF)
\textit{ by the nature of the institution and not solely on the basis of 
highest degree offered. Eligible PUIs are accredited colleges and 
universities (including two-year community colleges) that award Associate's
degrees, Bachelor's degrees, and/or Master's degrees in NSF- supported fields, 
but have awarded 20 or fewer Ph.D./D.Sci. degrees in all NSF-supported fields
during the combined previous two academic years.}\footnote{
\url{https://www.nsf.gov/funding/pgm\_summ.jsp?pim\_id=5518}} 
Simplifying slightly, PUIs can be thought of 
as teaching-focused institutions that do not offer Ph.D.s in Physics or 
Astronomy/Astrophysics. We are writing this White Paper primarily from the perspective 
and experience of faculty at 4-year Bachelor's degree-granting
institutions, which is heavily influenced by working in 
relatively small Physics/Astronomy Departments, where small typically 
means less than 10 faculty members and often means less than 5.

By a variety of measures PUIs make up the majority of the educational
opportunities in post-secondary education in the United States. 
In 2017, over $90\%$ of the 
$\sim$4,300 accredited post-secondary institutions were Non-Ph.D. granting 
institutions and over two-thirds of the nearly 20 million students enrolled 
in post-secondary education, did so at a non-Ph.D. granting institution.\footnote{
\url{https://nces.ed.gov/programs/digest/d18/tables/dt18\_317.40.asp}}
PUIs serve a broad range of student demographics, with many mission-driven PUIs focused 
on underserved populations. PUI enrollments are also heavily
weighted towards first-generation college students, particularly at 2-year 
institutions\footnote{\url{https://nces.ed.gov/pubs2018/2018009.pdf}}. 
Taking all of 
this together, PUIs offer an opportunity to dramatically influence 
the next generation of global citizens through the personalized student-faculty 
interactions that are often the hallmark of PUIs. This is true not only 
for STEM and astronomy-related 
careers, but for the general education of society as to the importance of 
literacy across a variety of intellectual, practical, and creative fields. 
As the astronomy community considers how to best support educating the next 
generation of astronomers it is important to appreciate the impact 
professional astronomers have beyond the astronomy community and how PUIs 
fit into this narrative.

Considering the current state of the profession, the astronomy community needs to do a 
better job recognizing PUIs as an integral part of the astronomy profession. This need is 
evidenced in the lack of PUI representation in the Decadal process itself. An analysis 
of membership 
on panels and committees from the 2010 Decadal suggests just 2 members out of 115 (1.4\%) 
were from PUIs, a number which vastly under-represents the contribution to astronomy research from faculty at PUIs. 

\vspace{0.25 cm}
\begin{tcolorbox}[left=0mm,right=0mm,width=6.5in,colback=gray!5!white,colframe=gray!75!black]
\justify
\textbf{Recommendation: The Decadal Survey panel and committee membership selection process 
should be adjusted to allow for greater involvement from faculty at PUIs. 
Solving this may require funds to enable teaching release for faculty serving on panels.}
\end{tcolorbox}\vspace{-0.25 cm}

\section{Astronomers and PUIs}\vspace{-0.25 cm}

At times in the past, and perhaps even continuing to this day, career 
paths leading to a position at a teaching-focused institution could 
be viewed by our professional community as a failure to secure a career 
at a research-focused institution, rather than the successful outcome 
it is. The reality
is many astronomers have intentionally made, and will continue to make, 
choices throughout their education and early-career years in an effort to 
realize a career at a PUI. For these astronomers, a career at a PUI is 
\textit{the goal}, not the consolation. The primarily 
teaching-focused job responsibilities of PUIs are not seen as a burden or 
at the expense of more research oriented careers, but rather as an opportunity to 
balance a genuine passion for education and mentoring with an ambitious 
research agenda designed to engage undergraduate students, while 
advancing the field of astronomy through original research and publication.

Clearly the reasons why an astronomer may choose a career at a PUI, or 
why a career at a PUI may be appropriate for an astronomer, can vary dramatically 
on a case-by-case basis. However, there are several appealing and distinct 
aspects associated with working at PUIs that are often highlighted in support 
of PUI careers. These reasons include 
i) a career where teaching and mentoring will be valued to a similar extent as
research productivity (and sometimes even more so), ii) the ability to mentor 
undergraduate students through substantial research projects, iii) the freedom 
to explore new research 
interests, including education and pedagogy research, iv) less pressure to secure
external funding. Interestingly, many astronomers at PUIs had their own 
positive educational experiences as students at PUIs
and the opportunity to 
provide similarly transformative experiences to the next generation of 
students is one of their primary motivators.

While it is common for astronomers to work at PUIs, stand-alone Astronomy 
Departments at PUIs are exceedingly rare and as such,
astronomers are primarily working in Physics Departments or 
more general Departments encompassing various natural science disciplines. 
It is not uncommon to be the only astronomer employed at the 
institution, resulting in opportunities and 
challenges that are somewhat unique in the professional astronomy community. 
Astronomers at PUIs are often expected to teach across the physics curriculum 
of their program, in addition to teaching or designing the department astronomy 
offerings and organizing campus or community-wide public observing events. 
These responsibilities make for an engaging and fulfilling career for 
astronomers at PUIs, but also underscore some of the training and experience 
necessary for an astronomer to be successful in the PUI environment, which is 
not necessarily the same training and experience necessary for success at a more
research-oriented institution. \vspace{-0.25 cm}

\section{Teaching}\vspace{-0.25 cm}

At nearly all PUIs, teaching is 
the primary job responsibility for faculty. What this means in practice can 
vary significantly across PUIs: at 2-year institutions, faculty often teach a 5-5 
load (meaning 5 courses in the fall semester and 5 courses again in the spring 
semester), while at 4-year institutions teaching loads can vary from 4-4 to 
2-2, where typically the lower teaching loads are balanced with expectations 
of greater grant and research productivity. For these reasons, it is difficult to 
generalize the typical teaching load for faculty working at PUIs. 
However, since teaching across the physics and astronomy curriculum is often 
a standard requirement at PUIs, an astronomer pursuing a career at a PUI
should be open to (and enjoy) teaching a variety of courses.

For many faculty working at PUIs, the emphasis on teaching is one of the 
most appealing aspects of the job. While there is a general expectation of 
teaching excellence, faculty at PUIs often have the freedom to explore
novel approaches to physics and astronomy pedagogy. This freedom is 
supported by the relatively small class sizes typically found at PUIs. 
This is particularly true for introductory astronomy courses where it 
is common for enrollments at PUIs to be an order of magnitude smaller 
than enrollments at many large research institutions. This situation 
allows faculty at PUIs to explore activities and assessments that 
may be more labor intensive, but which remain manageable with lower 
enrollments. An important fruit of smaller class sizes in introductory 
astronomy is the opportunity to engage students not 
majoring in quantitative fields with activities designed to deliberately 
encourage growth in quantitative literacy.  

While PUI faculty are encouraged to design and/or 
implement curriculum supported through various pedagogical research groups,
including both the Physics Education and Astronomy Education Research 
communities, faculty are also given the freedom to pursue 
approaches to curricular development that are often untenable in courses 
with large class sizes. This freedom can give rise to creative attempts 
to design labs and hands-on activities that address common misconceptions 
in physics and astronomy courses and can lead to peer-reviewed publications 
summarizing the design and implementation of the activity 
(e.g., Ribaudo 2016, 2017). 

When working at  a PUI it is common to teach introductory physics and astronomy 
courses, labs, general education courses, and advanced physics courses all within 
the relatively short time span of a few semesters. This variety demands 
faculty at PUIs remain intellectually flexible and adaptable throughout the 
course of their careers. PUIs also provide 
astronomers with freedom to explore pedagogy beyond the physical sciences. 
At many PUIs there is growing interest in multi- or interdisciplinary courses 
team-taught by faculty from different departments or divisions.  Thus, it is 
increasingly likely that astronomers at PUIs will be encouraged or required 
to develop courses with colleagues across the curriculum, including the arts, 
humanities, and social sciences.

Beyond having a general familiarity with broad physics topics and 
courses, it is helpful to have actual experience as the instructor 
of record when applying for faculty positions at PUIs. This poses
a challenge as the experience needed 
to land a permanent faculty position at many PUIs can most easily be 
obtained by working in a faculty position at a PUI. 
However, there are frequently visiting faculty positions available at PUIs 
that can serve as pedagogical ``post-docs".  Some are even formal post-docs 
with a faculty mentor and reduced teaching load designed specifically to help 
astronomers aspiring to a career at a PUI enhance their experience and skills.
One of our recommendations
is for the astronomy community to further support these types of post-doctoral
opportunities - allowing astronomers to better position themselves for a 
career at PUIs. 

\vspace{0.25 cm}
\begin{tcolorbox}[left=0mm,right=0mm,width=6.5in,colback=gray!5!white,colframe=gray!75!black]
\justify
\textbf{Recommendation: As a professional community we need to provide students and 
early-career astronomers with opportunities that will best position them for their careers. 
For those aspiring to a PUI-career this requires formal teaching experience, training, 
and mentoring. Ph.D. programs should expand their career training and support structures 
to include formal teaching opportunities. Funding agencies should support these efforts 
by establishing funding streams that allow for teaching and pedagogy-focused training.}
\end{tcolorbox}\vspace{-0.25 cm}

\section{Research and Mentoring}\vspace{-0.25 cm}

A common misconception regarding research at PUIs is that the faculty 
have little interest in establishing and maintaining an active research 
program. While it is true that the teaching responsibilities at PUIs can
make the process more challenging, nearly all astronomers at PUIs want to 
establish a sustainable research program. In fact, in all but the most extreme 
situations faculty at PUIs are expected to maintain an active scholarship 
portfolio. 

What an active scholarship profile translates to will depend on the specific institution, 
just as the teaching responsibilities at PUIs can also vary considerably
by institution. However, often the expectation is that PUI faculty are 
presenting their work at 
conferences, publishing their work in peer-reviewed journals, pursuing 
external funding, and perhaps most importantly, mentoring undergraduate 
student researchers and providing them the opportunity to experience the 
previously mentioned components of a scholarship portfolio. In other words, 
the scholarship portfolio at PUIs consist of many of the same 
components expected at more research-focused institutions. Of course, 
usually the frequency with which these components are expected to be realized 
is significantly lower than an astronomer would find at a more research-focused
institution, but some level of productivity is expected at nearly all 
PUIs.

A fundamental component to research programs at PUIs is the involvement of 
undergraduate students throughout the research process. In astronomy this can 
mean including students on observing proposals, bringing students on observing 
runs, training students to reduce and analyze various types of observational 
data, assisting students in the write-up of the project for publication, and 
arranging for students to present the results of the project at local, regional, 
or national conferences. For many students at PUIs, the research opportunities
provided by physics or astronomy faculty are the only opportunities these 
students will have to engage with an authentic research project. These 
experiences have the potential to be genuinely transformative for students, 
allowing for personal and professional growth well beyond what can be realized
in the traditional learning environment of the classroom. 

The important role faculty play in providing substantive research experiences 
cannot be overstated - in fact several recent studies have shown the majority of undergraduate 
research activities are not the result of sponsored programs from 
major national funding agencies (such as NSF, NASA, or NIH), but rather are the 
result of local or informal opportunities (Russell et al. 2007, Sadler et al. 2010). 
From Sadler et al. (2010), over 50\% of STEM majors engaged in independent or 
mentored research, while only 7\% of STEM majors were engaged with research sponsored 
by major national agencies. While faculty at PUIs are a critical component of the 
undergraduate research experience, the reality is the majority of PUI mentoring 
interactions are not formally or fully compensated by the institution (for faculty). 
This typically 
means the academic year research projects at PUIs are in addition to the already 
heavy teaching responsibilities and research and service expectations. Summer 
research projects can also be a challenge, with limited institutional support often 
requiring faculty at PUIs to mentor summer students without any compensation. 
Providing the appropriate support to allow for quality mentoring and research 
experiences at PUIs should be a priority for the astronomy community throughout the 
next decade.

Currently, there are just a few funding opportunities that specifically target 
PUI-affiliated participation and most of these are not designed exclusively for PUIs, 
but rather allow for the consideration of how a funding award may impact PUIs.
The most well-known of these in the astronomy community are awards associated 
with NSF programs, such as the Research at Undergraduate Institutions (RUI) and 
Research Opportunity Awards (ROA) programs. However the RUI program is only a 
designation that can be assigned to a proposal solicited from a separate program
and as such, RUIs are evaluated for intellectual merit and broader impacts 
along with all non-RUI proposals submitted. Additional NSF programs that often
fund or support PUI-affiliates include the Research Experiences for 
Undergraduates (REU), the Improving Undergraduate STEM Education (IUSE), and 
the New Faculty Workshop, however all of these programs solicit participation
from PUI and non-PUI applicants alike. Beyond the NSF, funding agencies such 
as NASA, NASA Space Grant, and Research Corporation regularly fund PIs from 
PUIs. But again, the majority of the available funding awarded by these agencies 
is not targeted to PUI-affiliates. 

\vspace{0.25 cm}
\begin{tcolorbox}[left=0mm,right=0mm,width=6.5in,colback=gray!5!white,colframe=gray!75!black]
\justify
\textbf{Recommendation: Funding agencies should increase the number of targeted 
grant programs for PUI-affiliates. Ideally this would include solicitations open only 
to PUI-affiliates and also include review-panels consisting primarily of PUI-affiliated
experts. Given the unique challenges associated with research at PUIs, allowing 
for awards up to 5 years would facilitate successful and sustainable projects. 
}
\end{tcolorbox}\vspace{-0.25 cm}

\section{Support and Resources Needed for the Next Decade}\vspace{-0.25 cm}

Over the last few decades the collective astronomy community has made a 
concerted effort to establish policies and best practices that allow for a 
more inclusive and supportive professional environment for astronomers 
from all backgrounds, identities, and career-trajectories. In doing so it 
has become clear there is a need within the community to  
better value the career opportunities accessible to astronomers that diverge 
from a research-focused career and value the 
contributions astronomers are making to the field from beyond research-focused
positions.\vspace{-0.25 cm} 

\subsection{Employment}

With regard to 
faculty employment opportunities, the American Institute of Physics 
(AIP) regularly reports employment statistics
for Physics and Astronomy Departments, providing a variety of metrics 
and graphics to characterize the employment environment across these 
adjacent fields. Several of these metrics establish that PUIs provide a 
substantive fraction of the full-time faculty positions in Physics and 
Astronomy.

\begin{figure}[h!]
    \begin{minipage}[c]{.5\textwidth}
      \centering
      \includegraphics[scale=.55]{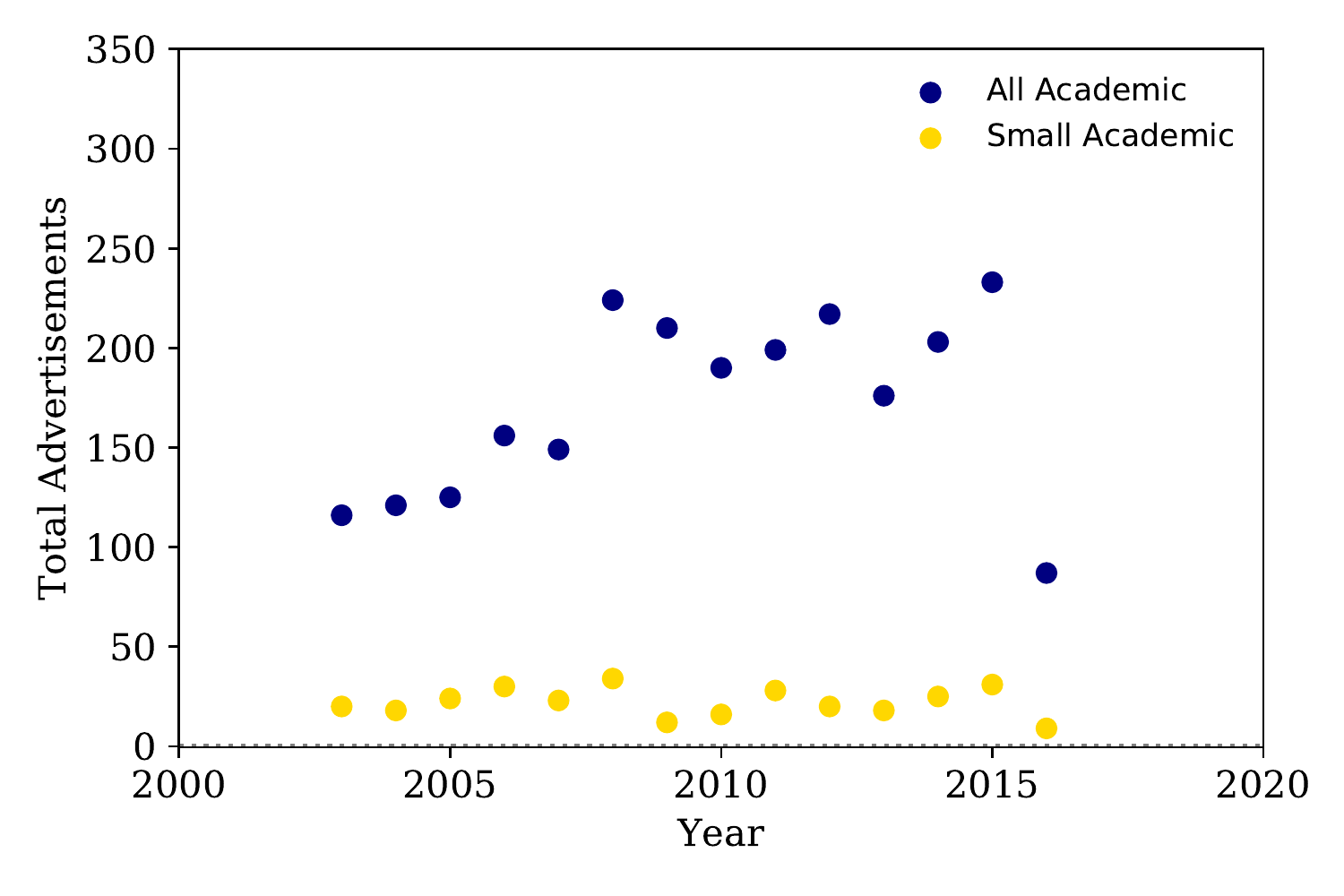}
    \end{minipage}
    \begin{minipage}[c]{.5\textwidth}
      \centering
      \includegraphics[scale=.55]{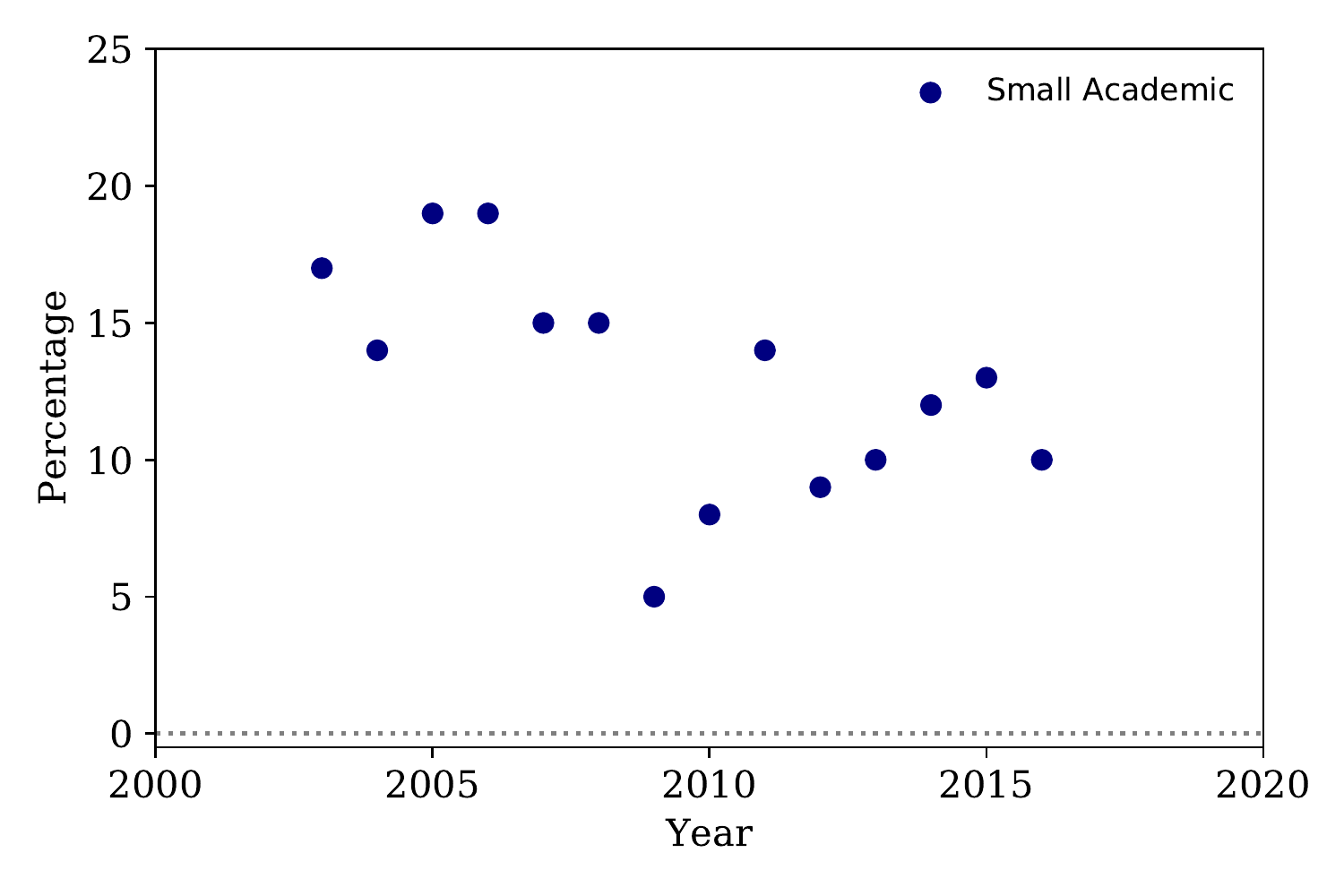}
    \end{minipage}\hfill
  
  \begin{minipage}[c]{\textwidth}
  \justify
    \caption{
    Shown here are the faculty job postings on the AAS Job Register over the last 
    15 years. \textbf{Left Panel:} the total number of academic faculty positions 
    advertised over the last 15 years along with the positions classified as `Small 
    Academic'. \textbf{Right Panel:} the percentage of faculty job postings classified  
    with institution type `Small Academic' over the last 15 years. For both figures 
    the positions include 
    both tenure-track and non tenure-track listings. 
    } \label{fig:jobs}
  \end{minipage}
\end{figure}

In an attempt to understand the employment opportunities at PUIs available to 
astronomers, we examined the archived job postings from the Job Register
maintained by the American Astronomical Society (AAS) and provided to us by the AAS 
Committee on Employment - see the White 
Paper by Kamenetzky et al. entitled \textit{Astronomy-driven Careers in 
the 2020's}. We performed a simple analysis of the archived job postings, which span roughly 
fifteen years, back to 2003. When a position is advertised on the Job Register, 
a variety of flags/categories get assigned - these include Institution Type, 
Position Type, Position Title, etc. The number of advertised faculty positions (tenure-track and 
non tenure-track) were examined for institutions classified as `Small Academic', the closest 
classification to the PUI distinction. Figure~\ref{fig:jobs} shows the results of our analysis. 
From this we see the relative number of astronomy 
positions at PUIs advertised on the Job Register has been gradually decreasing 
over the last decade. Taken at face value this would give the impression that
the opportunities for astronomers to pursue a career at PUIs are 
diminishing. However, the majority of openings at PUIs are in 
Physics Departments, often without a research specialty designation, meaning it is 
likely that the majority of openings at PUIs will not be advertised on 
the Job Register. 

\begin{figure}[h!]
    
    \begin{minipage}[c]{.5\textwidth}
      \centering
      \includegraphics[scale=.55]{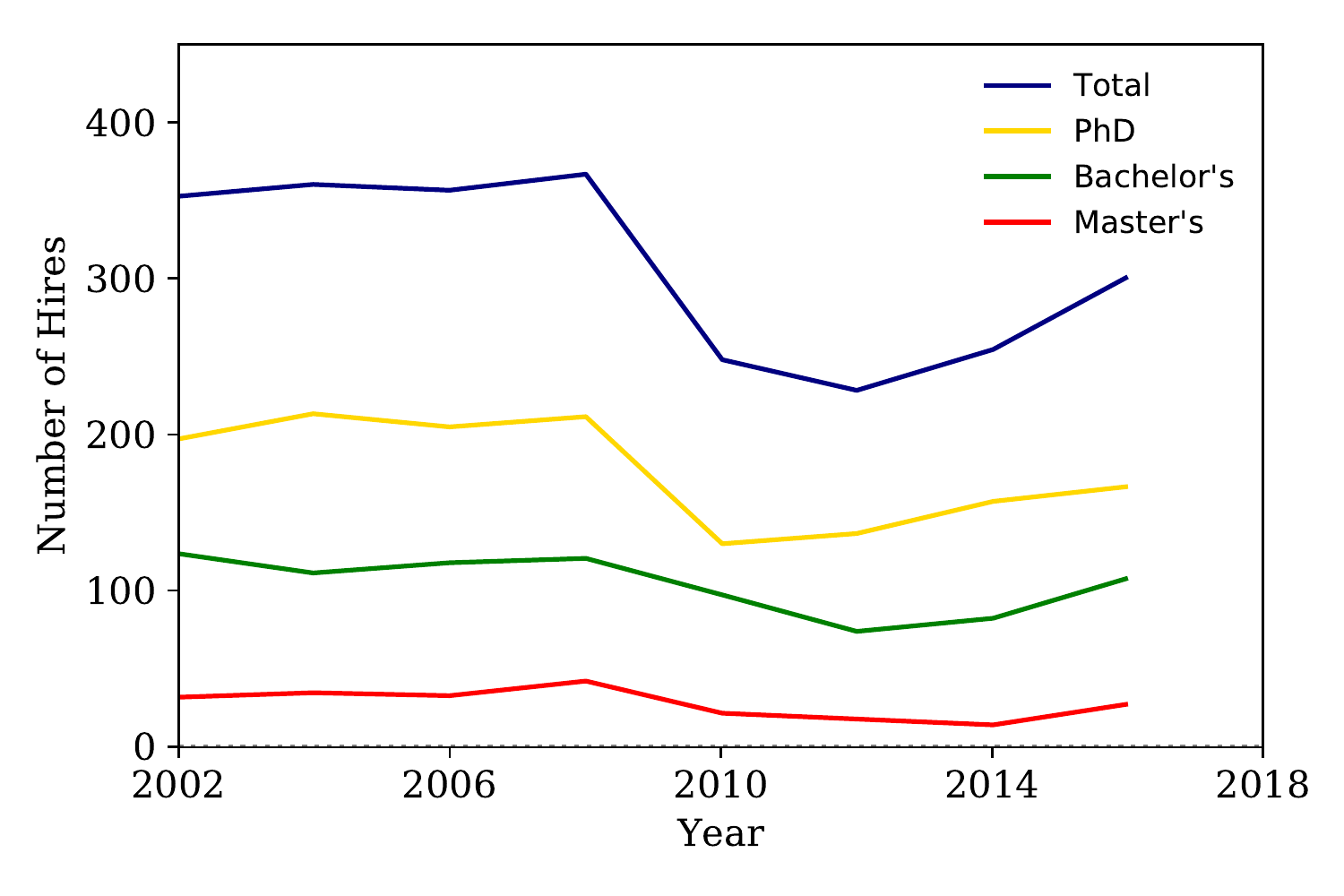}
    \end{minipage}
    \begin{minipage}[c]{.5\textwidth}
      \justify
    \caption{
    Shown here are the number of new faculty hires in Physics Departments as 
    reported by AIP in 2017. The increase in hires in recent years appears to be 
    driven by an increase in hiring at PUIs (both Bachelor's and Master's 
    degree-granting institutions). Note - this does not take into account
    the additional PUI hires at 2-year Associate's degree-granting institutions.
    } \label{fig:aip}
    \end{minipage}
    \hfill
  
\end{figure}

To expand on this, we examined several of the recent
AIP reports focused on employment at various institution types over the last
few decades.
Figure~\ref{fig:aip} is a recreation of the Fall 2017 
Physics Trends graphic published by the 
AIP,\footnote{\url{https://www.aip.org/statistics/physics-trends/number-faculty-hired-physics-departments}} where the number of tenured and 
tenure-track faculty hires in Physics Departments are plotted for 
various institution types over the last 15 years. Nearly half of the 
new Physics faculty hires in 2016 were hired into Departments at PUIs
(\uline{without taking into account the new hires at Associate's degree-granting
institutions}). This statistic is particularly important for the astronomy
community to be aware of as Ph.D.s with astronomy expertise
will be hired into Physics Departments at these institutions. 

\vspace{0.25 cm}
\begin{tcolorbox}[left=0mm,right=0mm,width=6.5in,colback=gray!5!white,colframe=gray!75!black]
\justify
\textbf{Recommendation: Expand the formal assessment of careers in astronomy 
to include an analysis of PUI-careers. This could be done as part of the AAS 
Committee on Employment or with the formation of a new AAS committee/sub-committee.
}
\end{tcolorbox}\vspace{-0.25 cm}

\vspace{-0.5 cm}

\vspace{-0.25 cm}

\subsection{Research Contributions}

With the potential for growth in PUI-careers over the last few decades, 
an interesting question to consider is whether the 
scholarly contributions of PUI-affiliates has evolved significantly 
over the same time frame. In what follows we have attempted to quantify the 
scholarship contributions of PUI-affiliates (faculty and students) to 
the astronomy community through 
an analysis of the publication and presentation archives accessible 
through the Astrophysics Database System (ADS) API.\footnote{\url{https://github.com/adsabs/adsabs-dev-api}}

Using the 2-year and 4-year institution list generated by the Council for 
Undergraduate 
Reserach,\footnote{\url{https://www.cur.org/assets/1/7/Slocum\_and\_Scholl-Table\_6.pdf}} 
we performed
searches of the ADS for entries published in each of The Astrophysical
Journal (ApJ - one of our most popular peer-reviewed publications) and the 
American Astronomical Society (AAS - one of our most popular venues for faculty and 
student presentation abstracts). 
While the CUR-lists are not identical 
to the NSF PUI distinction, we believe the significant overlap between the two 
lists allow for an overall assessment of the evolution of PUI-contributed 
research on the ADS. 

\begin{figure}[h!]
    \begin{minipage}[c]{.5\textwidth}
      \centering
      \includegraphics[scale=.55]{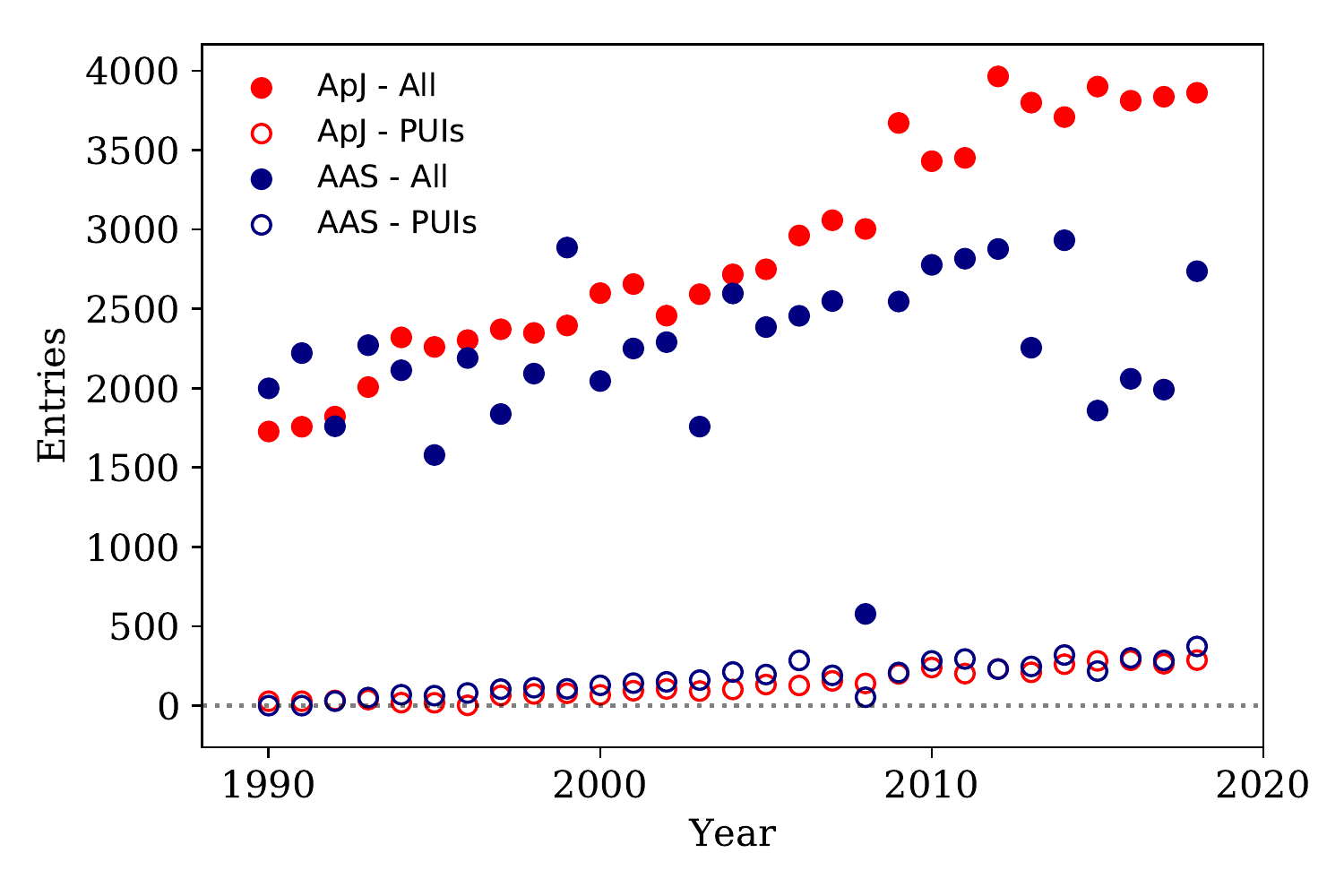}
    \end{minipage}
    \begin{minipage}[c]{.5\textwidth}
      \centering
      \includegraphics[scale=.55]{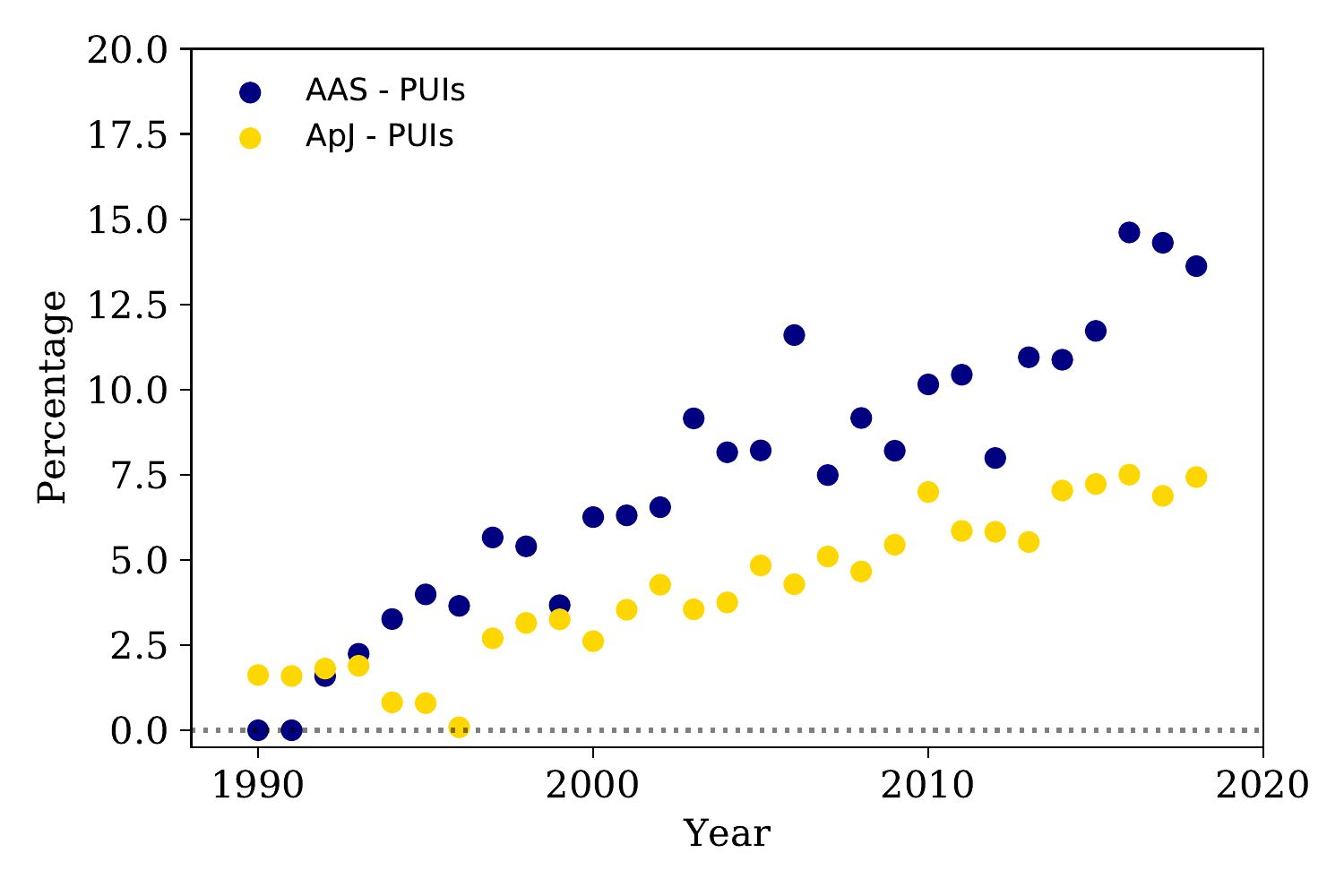}
    \end{minipage}\hfill
  
  \begin{minipage}[c]{\textwidth}
  \justify
    \caption{
    Shown here are the entries on the ADS with co-authors from PUIs
    (as matched through the ADS API with a PUI list from the Council on Undergraduate 
    Research). \textbf{Left Panel:} the total number of ADS entries associated 
    with ApJ and AAS, with entries from PUI-affiliated authors overplotted. \textbf{Right Panel:} 
    the percentage of the ApJ/AAS ADS entries with PUI-affiliated authors.
    } \label{fig:pubs}
  \end{minipage}
\end{figure}

Using a simple matching criterion we performed
total publication searches for PUI-affiliated authors relative to the total
number of publications for each year for the last thirty years. We find 
that the percentage of publications with PUI-affiliated authors has 
steadily increased for publications in ApJ and AAS. The general growth trends 
can be seen in Figure~\ref{fig:pubs}, where the right
panel shows the increase in AAS authorship is where the most 
significant growth in productivity has occurred. This isn't surprising as the 
AAS entries primarily correspond to presentation abstracts for the winter and 
summer AAS meetings, a typical venue for PUI faculty and 
undergraduate student presentations.

\vspace{-0.25 cm}

A more robust assessment of the scholarship contributions of PUI faculty and 
students would be appropriate for the astronomy community to prioritize in 
the next decade, allowing for a better understanding as to the direct impact 
PUI-affiliates are having on science productivity in astronomy. For example, 
the ApJ and AAS entries we examined as part of this work are only 
a portion of the publications cataloged with the ADS each year and it would 
be useful to analyze a larger portion of the ADS entries. A similar 
analysis should also be considered for the various funding agencies that 
typically support astronomy-related research.

\vspace{0.25 cm}
\begin{tcolorbox}[left=0mm,right=0mm,width=6.5in,colback=gray!5!white,colframe=gray!75!black]
\justify
\textbf{Recommendation: Establish a robust assessment of the contributions of 
PUI-affiliates when it comes to scholarly research 
publications, conference talks, and funding awards.
}
\end{tcolorbox}

\subsection{Regional Astronomy Groups}

It is not uncommon for an astronomer working at a PUI
to be the only astronomer at the institution. 
With this isolation comes the challenge of remaining engaged in the broader
astronomy community and maintaining an active research program with minimal
local support. One of the most 
important and accessible resources in this situation can be the regional astronomy 
community and regularly scheduled regional meetings. Regional meetings are 
particularly appealing when working at PUIs as the travel costs associated 
with attending are significantly lower than most national meetings. This 
is an important consideration because the financial support 
at many PUIs is not at a level that would cover travel costs for faculty 
and students to attend a national meeting. In addition, the timing of the 
January AAS meeting can lead to conflicts with the start of the academic 
term, making it a challenge for some PUI faculty and undergraduates to attend. 

For the majority of 
undergraduate students, the opportunity to simply attend a science meeting, let 
alone present an original research project, can have a dramatic impact on 
personal, professional, and social development. Regional meetings make 
these experiences more accessible for students at PUIs and are one 
of the primary reasons the astronomy community should make a deliberate 
effort in the next decade to support the development of more regional 
astronomy groups and their corresponding regional meetings. The blueprint 
for this already exists as regional 
meetings are ubiquitous throughout the physics community, with Sections 
of the American Physical Society (APS) and American Association of Physics
Teachers (AAPT) regularly organizing regional meetings.

The AAS officially recognizes and endorses regional meetings that 
satisfy a few basic rules and the meeting websites are linked-to on the 
AAS website.\footnote{\url{https://aas.org/meetings/regional-meetings}}
These regional meetings are not organized by the AAS, but rather 
a regional astronomy organization or a collective of astronomers from 
throughout a region. In the past five years the AAS has endorsed several 
regional meetings including the Mid-American Regional Astrophysics Conference, 
the Kentucky Area Regional Meeting, and the Meetings of the Astronomical 
Society of New York. 

For example, the Astronomical Society of New York (ASNY), has been 
organizing annual science conferences in the captial region of New York state
for the last 50 years. ASNY is the science arm of the New York Astronomical 
Corporation and there are more than 30 colleges and universities in New York 
that are currently member institutions affiliated with ASNY. ASNY serves as 
a critical resource for affiliates of PUIs and research institutions alike, 
providing student travel awards, 
student research prizes, and small-meeting opportunities for undergraduate 
and graduate students to present their research. PUI faculty benefit from 
all of these along with opportunities to take on significant service and 
leadership roles within the astronomy community. For these reasons, 
regional astronomy associations are an excellent investment for the astronomy
profession.

\vspace{0.25 cm}
\begin{tcolorbox}[left=0mm,right=0mm,width=6.5in,colback=gray!5!white,colframe=gray!75!black]
\justify
\textbf{Recommendation: Given the significant impact on the entire astronomy 
community, and particularly PUI-affiliates, the AAS and funding agencies should 
prioritize the development and support of regional astronomy associations along 
with regional astronomy meetings. Regional models used by the APS and AAPT 
should be used as guidance initially.
}
\end{tcolorbox}\vspace{-0.25 cm}

\subsection{PUI-focused Research Collaborations}

The essence of a research program at PUIs, where undergraduate student 
involvement is an integral component of the research experience, can lead to 
challenges that are unique to astronomers at PUIs. One of the 
primary issues is the fact that undergraduate research primarily consists 
of short-duration projects that 
are started and finished within a semester or a summer. Often students 
working on research projects have little course-work or training in the 
research field of their project and care must be taken to not overwhelm 
students early on in the project. In addition, the short duration of 
typical projects and the relatively little time a particular student 
will work on a project (often less than a year) make it difficult to 
establish a continuous and cohesive research strategy that seamlessly 
transitions from one student participant to the next. 
One possible remedy for this particular challenge is 
to build long-running research projects that benefit from the contributions 
of a large collaboration. 

Programs such as the SDSS FAST Initiative,\footnote{\url{https://www.sdss.org/education/faculty-and-student-team-fast-initiative/}} the Keck Northeast Astronomy Consortium (KNAC) REU,\footnote{\url{https://astro.swarthmore.edu/knac/}}
and the NSF-sponsored Undergraduate ALFALFA Team (UAT)\footnote{\url{http://egg.astro.cornell.edu/alfalfa/ugradteam/ugradteam.php}}
are specifically designed to engage 
PUI faculty and students with high-impact research experiences. These types of 
programs, where faculty are trained and exposed to research projects along with 
students are particularly effective at generating sustained research engagement. 
KNAC is an organization that falls both into the regional association 
category as well as a PUI-focused collaboration, consisting of a network of 
faculty and students from eight liberal arts colleges in the North East. The REU 
program pairs students with faculty for summer research projects and the fall 
KNAC Meeting allows for the students and faculty to share the results of their 
projects. With this Regional Consortium/REU model students and faculty are 
able to more easily maintain connections after the initial summer REU experience,
allowing for a more sustained impact compared to a typical REU.

Going beyond the REU model, the UAT was founded with faculty 
development as well as undergraduate research opportunities in mind, with the 
purpose to provide long-term collaborative research opportunities for faculty 
and students from a wide range of public and private undergraduate-focused
colleges in the context of the extragalactic ALFALFA HI survey.
In the past 12 years of funding,
34 mainly PUI faculty (44\% women) have participated. The UAT provides a support 
structure for PUI faculty that maximizes
the time that they can devote to the project and develops their research and education 
skills. Over the 12 years of funding, they 
have benefited from the opportunity to learn about and participate in the project, 
bringing their own sets of skills, experience, and expertise to the collaboration and 
developing a network of collaborators throughout the U.S. Further details are provided 
in \textit{Integrating Undergraduate Research and Faculty Development in a Legacy 
Astronomy Research Project} (Koopmann et al.).

\vspace{0.25 cm}
\begin{tcolorbox}[left=0mm,right=0mm,width=6.5in,colback=gray!5!white,colframe=gray!75!black]
\justify
\textbf{Recommendation: Funding agencies should prioritize and incentivize large-scale
research collaborations, including regional consortia, that primarily leverage 
PUI-affiliates to execute and support
the research objectives. These types of legacy research programs may not 
easily fit 
into the current funding program structures, highlighting the need for funding 
solicitations that encourage the development of new initiatives.
}
\end{tcolorbox}\vspace{-0.25 cm}

\pagebreak
\pagenumbering{gobble}

\section*{References}

\begin{itemize}[leftmargin=*,label={}]

\item Ribaudo, J.\ 2016, The Physics Teacher, 54, 330

\item Ribaudo, J.\ 2017, Physics Today, 70, 10

\item Ribaudo, Joseph; Koopmann, Rebecca A.; Haynes, Martha P.; Balonek, Thomas J.; Cannon, John M.; Coble, Kimberly A.; Craig, David W.; Denn, Grant R.; Durbala, Adriana; Finn, Rose; Hallenbeck, Gregory L.; Hoffman, G. Lyle; Lebron, Mayra E.; Miller, Brendan P.; Crone-Odekon, Mary; O'Donoghue, Aileen A.; Olowin, Ronald Paul; Pantoja, Carmen; Pisano, Daniel J.; Rosenberg, Jessica L. Troischt, Parker; Venkatesan, Aparna; Wilcots, Eric M.; ALFALFA Team 2017, AAS Meeting 229, id. 137.03 

\item Russell SH, Hancock MP, McCullough J. Benefits of undergraduate research experiences. Science. 2007;316:548–549

\item Sadler, Troy D., Burgin, S., McKinney,, L. 2010 ``Learning science through research apprenticeships: A critical review of the literature." Journal of Research in Science Teaching: The Official Journal of the National Association for Research in Science Teaching 47.3, 235

\end{itemize}

\theendnotes

\end{document}